\documentstyle[12pt]{article}

\catcode`\@=11

\ifx\documentclass\undefined
 \def\@sect#1#2#3#4#5#6[#7]#8{\ifnum #2>\c@secnumdepth
     \let\@svsec\@empty\else
     \refstepcounter{#1}\edef\@svsec{\csname prefix#1\endcsname
	\csname the#1\endcsname\hskip 1em}\fi
     \@tempskipa #5\relax
      \ifdim \@tempskipa>\z@
        \begingroup #6\relax
          \@hangfrom{\hskip #3\relax\@svsec}{\interlinepenalty \@M #8\par}%
        \endgroup
       \csname #1mark\endcsname{#7}\addcontentsline
         {toc}{#1}{\ifnum #2>\c@secnumdepth \else
                      \protect\numberline{\csname the#1\endcsname}\fi
                    #7}\else
        \def\@svsechd{#6\hskip #3\relax  
                   \@svsec #8\csname #1mark\endcsname
                      {#7}\addcontentsline
                           {toc}{#1}{\ifnum #2>\c@secnumdepth \else
                             \protect\numberline{\csname the#1\endcsname}\fi
                       #7}}\fi
     \@xsect{#5}}
\else
    \def\@seccntformat#1{\csname prefix#1\endcsname
	\csname the#1\endcsname\quad}
\fi

\@addtoreset{equation}{section}   

\def\thebibliography#1{\section*{References\@mkboth
 {REFERENCES}{REFERENCES}}\list
 {\leftbibmark\arabic{enumi}\rightbibmark}{
 \settowidth\labelwidth{\leftbibmark #1\rightbibmark}\leftmargin\labelwidth
 \advance\leftmargin\labelsep
 \usecounter{enumi}}
 \def\newblock{\hskip .11em plus .33em minus -.07em}
 \sloppy\clubpenalty4000\widowpenalty4000
 \sfcode`\.=1000\relax}

\def\@citex[#1]#2{\if@filesw\immediate\write\@auxout{\string\citation{#2}}\fi
  \def\@citea{}\@cite{\@for\@citeb:=#2\do
    {\@citea\def\@citea{,\penalty\@m\ }\@ifundefined
       {b@\@citeb}{{\bf ?}\@warning
       {Citation `\@citeb' on page \thepage \space undefined}}%
\hbox{\csname b@\@citeb\endcsname\citemarkdelim}}}{#1}}

\def\@cite#1#2{\leftcitemark{#1 \if@tempswa , #2\fi}\rightcitemark}
\def\leftcitemark{[}
\def\rightcitemark{]}
\def\citemarkdelim{}
\def\leftbibmark{[}
\def\rightbibmark{]}
\def\CITE#1{$^{\hbox{\tiny \cite{#1}}}$}

\catcode`\@=12

\def\L{{\cal L}}

\def\R{{\cal R}}

\def\Eq#1{Eq.(\ref{#1})}
\def\Eqs#1#2{Eqs.(\ref{#1})-(\ref{#2})}

\def\r#1{{\rm #1}}
\def\e#1{{10^{#1}}}
\def\bm#1{\mbox{\boldmath $#1$}}

\def\Frac(#1/#2){\left(\frac{#1}{#2}\right)}

\def\gsim{\stackrel{>}{\sim}}

\def\Tdot#1{{{#1}^{\hbox{.}}}}
\def\Tddot#1{{{#1}^{\hbox{..}}}}

\def\dddot#1{\stackrel{...}{#1}{}\!\!}
\def\Order#1{\r{O}\!\left(#1\right)}

\def\Beq{\begin{equation}}
\def\Eeq{\end{equation}}
\def\Beqr{\begin{eqnarray}}
\def\Eeqr{\end{eqnarray}}
\def\Beqrn{\begin{eqnarray*}}
\def\Eeqrn{\end{eqnarray*}}
\def\Bitm{\begin{itemize}}
\def\Eitm{\end{itemize}}

\font\elevenmib=cmmib10 scaled\magstephalf   \skewchar\elevenmib='177
\def\YUKAWAmark{\hbox{\elevenmib 
 Yukawa\hskip0.05cm Institute\hskip0.05cm Kyoto \hfill}}

\def\BP{ Z }

\begin{document}

\begin{titlepage}

\hbox to \hsize{\YUKAWAmark \hfill YITP-96-28}
\rightline{KUNS 1406}
\rightline{August 1996(Revised)}

\vspace{2cm}

\begin{center}\large\bf
Evolution of cosmological perturbations \\
in a stage dominated by an oscillatory scalar field
\end{center}

\bigskip

\begin{center}
Hideo Kodama\footnote{email address: kodama@yukawa.kyoto-u.ac.jp} 
\end{center}

\begin{center}\it
Yukawa Institute for Theoretical Physics, Kyoto University, \\
Kyoto 606-01, Japan\\
\end{center}

\begin{center}
and
\end{center}

\begin{center}
Takashi Hamazaki\footnote{email address: hamazaki@murasaki.scphys.kyoto-u.ac.jp}
\end{center}

\begin{center}\it
Department of Physics, Fuculty of Science, Kyoto University,\\
Kyoto 606-01, Japan
\end{center}

\bigskip
\bigskip
\begin{center}\bf Abstract\end{center}

In the investigation of the evolution of cosmological perturbations in
inflationary universe models the behavior of perturbations during the
reheating stage is the most unclear point. In particular in the early
reheating phase in which a rapidly oscillating scalar field dominates
the energy density, the behavior of perturbations is not known well
because their evolution equation expressed in terms of the curvature
perturbation becomes singular. In this paper it is shown that in spite
of this singular behavior of the evolution equation the Bardeen
parameter stays constant in a good accuracy during this stage for
superhorizon-scale perturbations except for a sequence of negligibly
short intervals around the zero points of the time derivative of the
scalar field.  This justifies the conventional formula relating the
amplitudes of quantum fluctuations during inflation and those of
adiabatic perturbations at horizon crossing in the Friedmann stage,
except for possible corrections produced by the energy transfer from
the scalar field to radiation in the late stage of reheating. It is
further shown that outside the above sequence of time intervals the
behavior of the perturbations coincides in a good accuracy with that
for a perfect fluid system obtained from the original scalar field
system by the WKB approximation and a spacetime averaging over a
Hubble horizon scale.

\end{titlepage}

\section{Introduction}

In the standard inflationary universe model the seeds of the present
large scale structures of the universe are generated from quantum
fluctuations of an inflaton field during
inflation\CITE{Hawking.S1982a,%
Guth.A&Pi1982,Starobinsky.A1982,Bardeen.J&Steinhardt&Turner1983}. Though
the amplitude and the spectrum of the seed fluctuations are easily
calculated when a model of the inflaton is given, it is not a simple
task to determined the amplitude and the spectrum of the corresponding
large scale perturbations when they reenter the Hubble horizon because
one must trace the evolution of the perturbations from the
inflationary stage to a recent time.

Of course it is not so hard to trace the evolution of perturbations
during the inflationary stage and the simple radiation-dominated
Friedmann stage after inflation. It is because the Bardeen parameter
$\BP$, which is defined as a linear combination of the curvature
perturbation $\Phi$ and the velocity perturbation $V$, is conserved
with a good accuracy for growing modes on superhorizon scales if the
equation of state of matter is regular and the entropy perturbation is
small\CITE{Bardeen.J&Steinhardt&Turner1983,Friemann.J&Turner1984,%
Brandenberger.R&Kahn1984,Kodama.H&Sasaki1984}. Hence, if the Bardeen
parameter is also conserved in the reheating stage which connects the
inflationary and the post-Friedmann stage, it is possible to determine
the amplitude and the spectrum of the perturbations at horizon
crossing just by the values of the Bardeen parameter for the seed
fluctuations at the inflationary stage. For example, it is the case in
the simplification that the reheating is instantaneous or no large
entropy perturbation is produced by reheating
processes\CITE{Friemann.J&Turner1984,%
Brandenberger.R&Kahn1984}. Actually in most of the work on the
cosmological perturbations in the inflationary scenario this
simplification is assumed. However, reheating is a rather complicated
stage in realistic models and entropy perturbations are not
necessarily negligible.  In such situations the conservation of
the Bardeen parameter for them is not well established.

For example, in the early stage of reheating, the energy density of
the universe is dominated by a rapidly oscillating inflaton field. In
this stage, if one neglects the tiny radiation contribution, the
equation of state of the inflaton shows an anomalous behavior and the
Bardeen parameter diverges periodically when the time derivative of
the inflation field vanishes. Accordingly the evolution equation for
the Bardeen parameter, or the closely related evolution equation for
the curvature perturbation becomes singular
periodically\CITE{Kodama.H&Sasaki1984}. Though this divergence is
replaced by periodic large peaks if the contribution of radiation is
taken into account and the height of the peaks decreases as the
inflaton decays, the coefficients of the evolution equations for $\BP$
and $\Phi$ still show bad behavior.  Hence it is not far from obvious
whether the values of the Bardeen parameter before and after the
reheating stage coincide with each other or not.

The main purpose of this paper is to show that in spite of this
peculiar behavior of the Bardeen parameter its value in the stage
dominated by a rapidly oscillating inflaton field actually coincides
with that in the preceding inflationary stage except for a sequence of
negligibly short intervals around the zero points of the time
derivative of the inflaton field, and that the behavior of the
curvature perturbation coincides with that for a perfect fluid system
obtained by the WKB approximation and a spacetime averaging from the
original scalar field system.  The main idea is to use the
gauge-invariant perturbation variable for the scalar field
perturbations introduced by Mukhanov in another
context\CITE{Mukhanov.V1988}, whose evolution equation is completely
regular and exactly solvable in the long-wavelength limit.

The paper is organized as follows. First in the next section the exact
solutions to the evolution equation for the Mukhanov variable in the
long-wavelength limit is given and their behavior is analyzed from the
point of view of the parametric resonance. Further by converting the
evolution equation into an integral equation with an iterative
structure with respect to the wavenumber, the behavior of the
solutions for general long-wavelength perturbations is described. The
singular nature of the transformation between the curvature
perturbation variable $\Phi$ and the Mukhanov variable in the
long-wavelength limit is also pointed out.  Then in \S3 the behavior
of the curvature perturbation and the Bardeen parameter is analyzed in
detail by solving the iterative integral equation for the Mukhanov
variable, first for the massive free scalar field and then for a
scalar field with a potential of a generic power-law type. In \S4 the
results obtained by this analysis is compared with the behavior of
perturbations for a perfect fluid system obtained by the WKB
approximation and a spacetime averaging from the original scalar field
system via a relativistic virial theorem.  Section 5 is devoted to
summary and discussion.

Throughout the paper the natural units $c=\hbar=1$ are adopted and
$8\pi G$ is denoted as $\kappa^2$. Further the notations for the
perturbation variables adopted in the article
\cite{Kodama.H&Sasaki1984} are used and their definitions are
sometimes omitted except for those newly defined in this paper

\section{Basic perturbation equations and their general analysis}

In this section we derive regular gauge-invariant equations for scalar
perturbations of a real scalar field and the gravitational field, and
give iterative expressions for their solutions.

In this paper we assume that the scalar field couples the
gravitational field minimally and its Lagrangian density is given by
\Beq
\L_\phi = -{1\over2}\sqrt{-g}\left(g^{\mu\nu}\partial_\mu\phi
\partial_\nu\phi + 2U(\phi)\right).
\Eeq
We will not specify the potential $U(\phi)$ in this section. 

Throughout the paper we only consider the case the unperturbed
background spacetime is described by a spatially flat Robertson-Walker
metric
\Beq
ds^2=-dt^2 + a^2 d\bm{x}^2.
\Eeq
Hence the unperturbed background of the scalar field follows the
equation
\Beq
\ddot \phi + 3H\dot\phi + U_\phi=0,
\label{EOM:phi:BG}\Eeq
and the scale factor $a$ is determined by 
\Beq
H^2:= \left(\dot a\over a\right)^2={\kappa^2\over6}
(\dot\phi^2+2U).
\label{HubbleEq}\Eeq
From these equations the time derivative of $H$ is given by
\Beq
\dot H=-{\kappa^2\over2}\dot\phi^2.
\label{Hdot}\Eeq
Throughout the paper it is understood that all the perturbation
variables represent Fourier expansion coefficients of perturbations of
the corresponding quantities around this background. For simplicity we
omit the wave number suffix $k$ for them.

\subsection{$(X,\Phi)$-system}

The most important geometrical quantity describing scalar
perturbations is given by the gauge-invariant variable $\Phi$ defined
by
\Beq
\Phi:=\R - {aH\over k}\sigma_g,
\Eeq
where $\R$ and $\sigma_g$ are the perturbations of the three curvature
and the shear of each constant time slice,
respectively\CITE{Bardeen.J1980,%
Kodama.H&Sasaki1984}.  This quantity
represents the curvature perturbation of the Newtonian slice
$\sigma_g=0$. Therefore a natural gauge-invariant variable constructed
from the perturbation $\delta\phi$ of the scalar field is
\Beq
X:=\delta\phi -{a\over k}\dot\phi \sigma_g.
\Eeq

From the components $(0,0), (0,j)$ and the traceless part of $(j,k)$
of the perturbed Einstein equations, we obtain the following evolution
equations for $X$ and $\Phi$\CITE{Kodama.H&Sasaki1984}:
\Beqr
&&\dot X\dot\phi -\ddot\phi X +\dot\phi^2\Phi={2\over\kappa^2}{k^2\over a^2}
\Phi,\label{Hconstraint:X-Phi}\\
&&\dot\Phi + H\Phi =-{\kappa^2\over 2}\dot\phi X.
\label{Mconstraint:X-Phi}\Eeqr
These equations correspond to the Hamiltonian and the momentum
constraints, respectively. The equations obtained from the other
components do not give new equations independent from these.

On the other hand the perturbed field equation for $\phi$ 
gives\CITE{Kodama.H&Sasaki1984}
\Beq
\ddot X + 3H\dot X + \left({k^2\over a^2}+U_{\phi\phi}\right)X
=-4\dot\phi\dot\Phi + 2U_\phi\Phi.
\label{FieldEq:X}\Eeq
This equation is also automatically satisfied under
\Eqs{Hconstraint:X-Phi} {Mconstraint:X-Phi} due to the Bianchi
identity if we require smoothness of $X$ and $\Phi$.

By eliminating $\Phi$ from \Eq{FieldEq:X} with the help of
\Eqs{Hconstraint:X-Phi}{Mconstraint:X-Phi}, we obtain a closed 
2nd-order equation for $X$:
\Beq
\ddot X + 3H\dot X + \left({k^2\over a^2}+U_{\phi\phi}
-2\kappa^2\dot\phi^2\right)X
={2(2H\dot\phi+ U_\phi)\over\dot\phi^2-{2\over\kappa^2}{k^2\over a^2}}
(\ddot\phi X -\dot \phi \dot X).
\Eeq
Similarly by eliminating $X$, we obtain an equation for $\Phi$:
\Beq
\Tdot{\left({1\over a\dot\phi^2}\Tdot{(a\Phi)}\right)}
+\left({k^2\over a^2\dot\phi^2}-{\kappa^2\over2}\right)\Phi=0.
\Eeq
Both of these equations are singular. Hence, though the independent
degree of freedom is two in the current system, it is difficult to
analyze the behavior of perturbations by a 2nd-order differential
equation in terms of $X$ and $\Phi$. In particular this singular
nature make it impossible to construct solutions iteratively with
respect to $k$, as we will see later.

In spite of this difficulty the above set of equations has one good
point.  It is that we can easily solve them in the long-wavelength
limit $k=0$.  To see this, let us rescale $\Phi$ as
\Beq
u:={a\over H}\Phi.
\Eeq
Then for $k=0$, \Eqs{Hconstraint:X-Phi}{Mconstraint:X-Phi} are written
as
\Beqr
&& \Tdot{\left(X\over\dot\phi\right)}=-{H\over a}u,\\
&&\dot u-{\kappa^2\over 2H}\dot\phi^2 u
=-{\kappa^2\over2}{a\dot\phi\over H}X.
\Eeqr
By eliminating $X$ from these equations we obtain
\Beq
\Tdot{\left({H\over a\dot\phi^2}\dot u\right)}={\kappa^2\over 2a}\dot u.
\Eeq
This equation can be easily solved to give
\Beq
u=A-{\kappa^2\over 2}B\int _{t_0}{a\dot\phi^2\over H^2}dt,
\Eeq
where $A$ and $B$ are integration constants, and $t_0$ is some initial
time. Hence for $k=0$ the exact solutions for $(X,\Phi)$-system are
given by
\Beqr
&&\Phi=A{H\over a}-B{\kappa^2\over 2}{H\over a}
\int _{t_0}{a\dot\phi^2\over H^2}dt,
\label{Sol:Phi:X-Phi}\\
&& X=A{\dot\phi\over a} + B\left({\dot\phi\over H}
+{\dot\phi\over a}\int _{t_0}{a\dot\phi^2\over H^2}dt\right).
\label{Sol:X:X-Phi}\Eeqr
These solutions are completely regular and will play an important role
in the following analysis.

With the help of these solutions and the Green function method we can
transform the 2nd-order equation for $\Phi$ above to the following
iterative integral equation:
\Beq
\Phi= AU_1 + BU_2 + 2k^2U_1 \int _{t_0} {U_2\over a^2\dot\phi^2}\Phi dt
-2k^2U_2 \int _{t_0} {U_1\over a^2\dot\phi^2}\Phi dt,
\Eeq
where 
\Beqr
&& U_1:={H\over a},\\
&& U_2:={H\over a}\int_{t_0}{a\dot\phi^2\over H^2}dt
={2\over\kappa^2}\left(1-{H\over a}\int_{t_0}a dt\right).
\Eeqr
It is clear that this integral equation yields an expression which is
singular at $\dot\phi=0$ when one tries to solve it iteratively.
Hence the system $(X,\Phi)$ is not appropriate for investigating the
long-wavelength behavior of the perturbations as commented above.

\subsection{$Y$-system}

The above difficulty in the $(X,\Phi)$-system is resolved if we use a
new gauge-invariant variable introduced by
Mukhanov\CITE{Mukhanov.V1988,Makino.N&Sasaki1991}. It is defined by
\Beq
Y:=X-{\dot\phi\over H}\Phi.
\label{Y:def}\Eeq
Since it is expressed as
\Beq
Y=\delta\phi -{\dot\phi\over H}\R,
\Eeq
it represents the perturbation amplitude of the scalar field on the 
flat slices $\delta\R=0$. 

This quantity has also a close relation to the Bardeen
parameter\CITE{Bardeen.J&Steinhardt&Turner1983} defined by
\Beq
\zeta:=\left(1+{2\over 9}{k^2\over (1+w)H^2a^2}\right)\Phi -{aH\over k}V,
\label{BardeenParam:def}\Eeq
where $w=P/\rho$, and $V$ is a gauge-invariant quantity representing
the velocity perturbation in the isotropic gauge. This gauge-invariant
quantity represents the curvature perturbation of uniform-Hubble
slices\CITE{Bardeen.J1980} and is conserved approximately for
superhorizon perturbations if $1+w$ is a positive quantity of order
unity. In this paper instead of this original Bardeen parameter we use
the gauge-invariant variable defined by
\Beq
\BP:=\Phi -{aH\over k}V,
\Eeq
and refer it as the Bardeen parameter because it coincides with the
original one for superhorizon perturbations under the same condition
on $1+w$ as above, and shares the same good
property\CITE{Kodama.H&Sasaki1984}.

Further it is more suitable for the analysis of the problem we are
concerned. In particular in the present case in which matter consists
only of the scalar field $V$ is expressed in terms of $X$
as\CITE{Kodama.H&Sasaki1984}
\Beq
V={k\over a\dot\phi}X.
\Eeq
Hence $\BP$ is simply written in terms of $Y$ as
\Beq
\BP=-{H\over\dot\phi}Y.
\label{BardeenParamByY}\Eeq

By a simple calculation it can be shown that 
\Eqs{Hconstraint:X-Phi}{Mconstraint:X-Phi} leads to the regular
closed 2nd-order differential equation of $Y$,
\Beq
\ddot Y + 3H\dot Y + \left({k^2\over a^2}+U_{\phi\phi}+
3\kappa^2\dot\phi^2-{\kappa^4\over 2H^2}\dot\phi^4
+2\kappa^2{\dot\phi\over H}U_\phi\right)Y=0.
\label{Eq:Y:general}
\Eeq
This equation is equivalent to the original system
(\ref{Hconstraint:X-Phi}) and (\ref{Mconstraint:X-Phi}) for $k\not=0$
because by the inverse transform from $Y$ to $\Phi$ and $X$ given by
\Beqr
&&\Phi={\kappa^2\over2}{a^2 H\over k^2}\left[{\dot\phi \over H}\dot Y
-\Tdot{\left(\dot\phi\over H\right)}Y\right],
\label{PhiByY}\\
&& X= Y + {\dot\phi\over H}\Phi
\label{XByY}
\Eeqr
we can recover the original equations for $X$ and $\Phi$.

In the limit case $k=0$, however, this inverse transform is singular.
Hence if we take the limit by keeping $\Phi$ finite, we obtain the
additional constraint on $Y$ given by
\Beq
\dot\phi \dot Y - H\Tdot{\left(\dot\phi\over H\right)}Y=0.
\Eeq
Due to this the exact solutions (\ref{Sol:Phi:X-Phi}) and 
(\ref{Sol:X:X-Phi}) for the $X$ and $\Phi$ in the limit $k=0$
reduce to the single solution for $Y$,
\Beq
Y=B{\dot\phi\over H}.
\Eeq
It is easily confirmed by a direct calculation that this satisfies
\Eq{Eq:Y:general} for $k=0$. Of course all the solutions for $k=0$ 
can be constructed from this special solution with the help of the
standard procedure as
\Beqr
&& Y=A V_1 + B V_2;\\
&& V_1:={\dot\phi\over H},
\label{V1:def}\\
&& V_2:={\dot\phi\over H}\int_{t_0}{H^2\over a^3\dot\phi^2}dt.
\label{V2:def}\Eeqr

The new independent solution $V_2$ is apparently singular at
$\dot\phi=0$. However, it is completely regular actually as is
expected from the regularity of the original equation for $Y$. To be
more precise, it has a finite limit at $\dot\phi=0$ and can be
uniquely extended across that point smoothly.  In order to find an
explicit expression for this extended solution $V_2$, let us denote
the sequence of time at which $\dot\phi=0$ by $\tau_j$, and examine
the behavior of $V_2$ around $t=\tau_j$.

If we take a sequence of time $t_j$ so that $\tau_{j-1}< t_j <
\tau_j$, then $V_2$ should be expressed in each interval $\tau_{j-1} < 
t < \tau_j$ as
\Beq
V_2=A_j V_1 + B_j {\dot\phi\over H}\int_{t_j}{H^2\over a^3\dot\phi^2}dt,
\label{V2:sequence}\Eeq
where $A_0=0$ and $B_0=1$. In order to find the recurrence relations
for $A_j$ and $B_j$, let us expand $\phi$ at $t=\tau_j$ as
\Beq
\phi=\phi_j + {1\over 2}\ddot\phi_j(t-\tau_j)^2
+{1\over 6}\dddot\phi_j(t-\tau_j)^3+ \cdots,
\Eeq
where the suffix $j$ represents the value at $t=\tau_j$. Inserting
this expansion into \Eq{EOM:phi:BG} and \Eq{HubbleEq}, we obtain
\Beqr
&& \ddot\phi_j=-U_{\phi j},\quad \dddot\phi_j=-3H_j \ddot\phi_j,\\
&&H=H_j\left[1+\Order{(t-\tau_j)^3}\right],\\
&&a=a_j\left[1+H_j(t-\tau_j)+\Order{(t-\tau_j)^2}\right].
\Eeqr
Hence the integrand of $V_2$ is expressed as
\Beq
{H^2\over \dot\phi^2 a^3}={H_j^2\over \ddot\phi_j^2 a_j^3}
\left[{1\over (t-\tau_j)^2}+\Order{1}\right].
\Eeq

From this it follows that in the interval $\tau_{j-1}<t<\tau_j$ the
integration in \Eq{V2:sequence} is written as
\Beq
\int_{t_j} {H^2\over \dot\phi^2 a^3}dt
=\int_{t_j}\left[{H^2\over \dot\phi^2 a^3}
-{H_j^2\over \ddot\phi_j^2 a_j^3}{1\over (t-\tau_j)^2}\right]dt
+{H_j^2\over \ddot\phi_j^2 a_j^3}\left({1\over \tau_j-t}
-{1\over \tau_j-t_j}\right),
\Eeq
and in the interval $\tau_j < t < \tau_{j+1}$ as
\Beq
\int_{t_{j+1}} {H^2\over \dot\phi^2 a^3}dt
=\int_{t_{j+1}}\left[{H^2\over \dot\phi^2 a^3}
-{H_j^2\over \ddot\phi_j^2 a_j^3}{1\over (t-\tau_j)^2}\right]dt
+{H_j^2\over \ddot\phi_j^2 a_j^3}\left({1\over \tau_j-t}
-{1\over \tau_j-t_{j+1}}\right).
\Eeq
Since the integrands in the right-hand sides of these equations are
regular at $t=\tau_j$, $V_2$ has the finite limits at $t=\tau_j\pm0$
given by
\Beqr
&&V_2(\tau_j-0)=-B_j{H_j\over \ddot\phi_j a_j^3},\\
&&V_2(\tau_j+0)=-B_{j+1}{H_j\over \ddot\phi_j a_j^3}.
\Eeqr
Hence from the continuity of $V_2$ at $t=\tau_j$ we obtain
\Beq
B_j=B_0=1.
\Eeq

Similarly $\dot V_2$ also has the finite limits given by
\Beqr
&&\dot V_2(\tau_j-0)=A_j{\ddot \phi_j\over H_j}
 + {H_j\over \ddot\phi_j a_j^3}\left({3\over 2}H_j
-{1\over \tau_j -t_j}\right)
+{\ddot\phi_j\over H_j}\int_{t_j}^{\tau_j}\left[{H^2\over \dot\phi^2 a^3}
-{H_j^2\over \ddot\phi_j^2 a_j^3}{1\over (t-\tau_j)^2}\right]dt,
\nonumber\\&&\\
&&\dot V_2(\tau_j+0)=A_{j+1}{\ddot \phi_j\over H_j}
 + {H_j\over \ddot\phi_j a_j^3}\left({3\over 2}H_j
-{1\over \tau_j -t_{j+1}}\right) 
 + {\ddot\phi_j\over H_j}\int_{t_{j+1}}^{\tau_j}\left[{H^2\over \dot\phi^2 a^3}
-{H_j^2\over \ddot\phi_j^2 a_j^3}{1\over (t-\tau_j)^2}\right]dt.
\nonumber\\&&
\Eeqr
Hence the continuity of $\dot V_2$ gives the recurrence relation
\Beq
A_{j+1}-A_j =
\int_{t_j}^{t_{j+1}}\left[{H^2\over \dot\phi^2 a^3}
-{H_j^2\over \ddot\phi_j^2 a_j^3}{1\over (t-\tau_j)^2}\right]dt
  -{H_j^2\over a_j^3\ddot\phi_j^2}
\left({1\over \tau_j-t_j} +{1\over t_{j+1}-\tau_j}\right).
\label{RecurrenceRelationForA}\Eeq
In the next section we will show that we can take $t_j$ so that it is
near the zero points of $\phi$ and $A_j=0$.

This argument shows that $V_2$ takes a definite value independent of
$t_0$ at $t=\tau_j$,
\Beq
V_2(\tau_j)=-{H(\tau_j)\over a(\tau_j)^3\ddot\phi(\tau_j)},
\label{V2(tau)}\Eeq
which changes its sign alternatively with $j$. We will show in the
next section that it approximately gives the amplitude of $V_2$ at
each interval and its absolute value decreases monotonically with $j$. 
Hence $V_2$ is a decaying mode. On the other hand from \Eq{HubbleEq}
the amplitude of $V_1$ stays nearly constant. This behavior should be
contrasted with that of $\phi$ whose amplitude decreases monotonically
in the rapidly oscillating phase, and can be understood as a result of
the parametric resonance.

For example, for the case $U=m^2\phi^2/2$, the equation for $Y$
is written as 
\Beq
\Tddot{(a^{3/2}Y)}+ \left[m^2+{4m\over t}\sin 2m(t-t_0)\right]
(a^{3/2}Y)\simeq0,
\Eeq
under the condition $(k^2/a^2)/mH\ll1$ and $mt\gg1$(see
\Eq{EOM:phi:BG'}). This equation satisfies the condition for the
parametric resonance to occur. Since it can be put into the integral
equation
\Beqr
&&a^{3/2}Y= A\cos m(t-t_0) + B\sin m(t-t_0) \nonumber\\
&& \qquad +\cos m(t-t_0)\int_{t_0} dt \sin m(t-t_0)
{4\over t} \sin 2m(t-t_0) a^{3/2}Y \nonumber\\
&& \qquad -\sin m(t-t_0)\int_{t_0} dt \cos m(t-t_0)
{4\over t} \sin 2m(t-t_0) a^{3/2}Y,
\Eeqr
the amplitude of the solution corresponding to $A=1$ and $B=0$ is
multiplied every each period by
\Beq
1+\int_{t_0}^{t_0+\Delta t} dt {2\over t}\sin^2 2m(t-t_0)
\simeq 1+{\Delta t\over t},
\Eeq
where $\Delta t$ is the period of oscillation. Hence after a long time
$a^{3/2}Y$ behaves as
\Beq
a^{3/2}Y\simeq e^{\int_{t_0}{dt\over t}}\cos m(t-t_0)
={t\over t_0}\cos m(t-t_0).
\Eeq
Similarly for the solution corresponding to $A=0$ and $B=1$ the
amplitude is multiplied by $1-\Delta t/t$ every each period and
behaves as
\Beq
a^{3/2}Y\simeq e^{-\int_{t_0}{dt\over t}}\cos m(t-t_0)
={t_0\over t}\cos m(t-t_0).
\Eeq
It is easy to see that this behavior coincides with that of $V_1$ and
$V_2$ approximately.

On the other hand  \Eq{EOM:phi:BG} is written as
\Beq
\Tddot{(a^{3/2}\phi)}+\left[m^2+\Order{1\over t^2}\right](a^{3/2}\phi)=0.
\Eeq
In this case, since $\int^\infty_{t_0} dt/(mt^2)$ converges to a finite
quantity of order $1/(mt_0)$, no parametric resonance occurs for
$mt_0\gsim1$.

Here note that the amplitude of $Y$ increases in proportion to a power
of time by the effect of parametric resonance in contrast to the
Mathieu function in which the amplitude grows exponentially. This
feature is common to the parametric resonance of coupled oscillating
fields on expanding universes whose scale factor increases in
proportion to a power of time as the above argument
shows(cf. Ref.\cite{Shatanov.Y&Traschen&Brandenberger1995}).

\subsection{Iterative solutions}

When the wave number $k$ is small, we can convert the differential
equation (\ref{Eq:Y:general}) into an iterative integral equation with
the help of the solutions $V_1$ and $V_2$ for $k=0$ and the Green
function method:
\Beqr
&Y
&=A V_1 +B V_2 + k^2V_1\int_{t_0} aV_2 Y dt 
-k^2 V_2\int_{t_0} aV_1 Y dt \nonumber \\
&&=A V_1 + B V_2 + k^2 G\circ (aY).
\label{IterativeEq:Y}\Eeqr
From this equation and \Eq{PhiByY} $\Phi$ is expressed as
\Beq
\Phi={\kappa^2\over2}{H\over ak^2}W(V_1,Y),
\label{PhiByYandV1}\Eeq
where $W$ is the Wronskian of \Eq{Eq:Y:general} given by
\Beq
W(u,v):=a^3(u\dot v-v \dot u).
\Eeq

Let $Y_1$ and $Y_2$ be the solutions to the above integral equation
for $(A,B)=(1,0)$ and $(A,B)=(0,1)$, respectively, and $\Phi_1$ and
$\Phi_2$ be the corresponding solutions for $\Phi$. Then since
\Beq
W(Y_1,Y_2)=W(V_1,V_2)=1,
\Eeq
from \Eq{PhiByYandV1} $\Phi_1$ and $\Phi_2$ are written as
\Beqr
&&\Phi_1=-{\kappa^2\over2}{H\over a}\int_{t_0}aV_1 Y_1 dt,
\label{Phi1ByY}\\
&&\Phi_2={\kappa^2\over2}{H\over ak^2}
\left(1-k^2\int_{t_0}aV_1 Y_2 dt\right).
\label{Phi2ByY}\Eeqr
We will show in the next section that $Y_1\simeq V_1$ and $Y_2\simeq
V_2$, and the second term on the right-hand side of \Eq{Phi2ByY} is
much smaller than unity for superhorizon scale perturbations. From
this it follows that
\Beqr
&& \Phi_1 \simeq \r{const},\\
&& \Phi_2 \simeq {\kappa^2\over 2}{H\over ak^2}.
\Eeqr
Hence in spite of the singularity of the transformation (\ref{PhiByY})
and (\ref{XByY}) the solutions (\ref{Sol:Phi:X-Phi}) and
(\ref{Sol:X:X-Phi}) for the $(X,\Phi)$-system with $k=0$ yield a good
approximation for the behavior of superhorizon scale perturbations if
we give appropriate $k$-dependences to the constants $A$ and $B$.

\section{Behavior of the curvature perturbation and the
Bardeen parameter}

In this section we solve the integral equation for $Y$ given in the
previous section and analyze the behavior of the curvature
perturbation and the Bardeen parameter for superhorizon perturbations
in the oscillatory phase of the scalar field. First we look at the
case $U$ is proportional to $\phi^2$ in detail and then extend the
analysis to the generic case $U\propto |\phi|^n$.

\subsection{$U=m^2\phi^2/2$ case}

In the case in which the potential of the scalar field is given by
\Beq
U={1\over2}m^2\phi^2,
\Eeq
\Eq{EOM:phi:BG} is rewritten as
\Beq
\Tddot{(a^{3/2}\phi)} 
+ \left[m^2+{3\over2}\left(\dot H + {3\over2}H^2\right)\right]
(a^{3/2}\phi)=0.
\label{EOM:phi:BG'}\Eeq
From Eqs.(\ref{HubbleEq}) and (\ref{Hdot}) the solution to this
equation is estimated as
\Beqr
&a^{3/2}\phi &=\sigma \sin m(t-t_0)+{3\over 2m}\sin mt \int_{t_0}
\left(\dot H+{3\over2}H^2\right)\cos mt \,a^{3/2}\phi dt \nonumber\\
&& \qquad -{3\over 2m}\cos mt \int_{t_0}
\left(\dot H+{3\over2}H^2\right)\sin mt \,a^{3/2}\phi dt \nonumber\\
&& = \sigma\left[\sin m(t-t_0)
-{1\over 4m}\left({1\over t_0}-{1\over t}\right)\cos m(t-t_0) 
+ \Order{\epsilon^2}\right],
\label{phi:harmonic}\Eeqr
where
\Beq
\epsilon:= {H\over m}.
\Eeq
Here note that in the oscillatory phase of $\phi$, $\epsilon\ll1$.
Similarly $\dot\phi$ is estimated as
\Beq
a^{3/2}\dot\phi=m\sigma\left[\cos m(t-t_0)
+{1\over 4m}\left({1\over t_0}-{5\over t}\right)\sin m(t-t_0) 
 + \Order{\epsilon^2}\right].
\label{phidot:harmonic}\Eeq
By inserting these expressions into \Eq{HubbleEq} and solving it we
obtain
\Beqr
&& H={2\over3t}\left[1-{1\over2mt}\sin 2m(t-t_0)+\Order{\epsilon^2}\right],
\label{H:harmonic}\\
&& a^3={3\over8}\kappa^2\sigma^2(mt)^2\left[1+\Order{\epsilon^2}\right].
\label{a:harmonic}\Eeqr

Next we determine the explicit behavior of $V_1$ and $V_2$ with the
help of these equations. Since we need an estimate for
$a^3\dot\phi^2/H^2$ which is correct up to the order $(t-\tau_j)^3$ at
each interval $t_j<t<t_{j+1}$, we first estimate its second derivative
and then integrate it.

Let us denote the sequence of the zero points of $\dot\phi$ by
$\tau_j$ and take another sequence of time, $t_j$, as in the previous
section. Then from \Eq{phidot:harmonic} the quantity $\xi_j$ defined
by
\Beq
\xi_j:=m(\tau_j-t_0),
\Eeq
satisfies the equation
\Beq
\cos \xi_j=\left({5\over 4 m\tau_j}-{1\over 4mt_0}\right)\sin\xi_j 
+ \Order{\epsilon_j^2},
\label{xi:estimate}\Eeq
where $\epsilon_j=\epsilon(\tau_j)$. This gives the estimate
\Beq
\sin\xi_j=(-1)^j + \Order{\epsilon_0^2}.
\Eeq
Hence in the interval $t_j<t<t_{j+1}$, $\Tddot{(a^{3/2}\dot\phi/H)}$
is estimated as
\Beqr
&\Tddot{\left(a^{3/2}\dot\phi\over H\right)}
&= -{m^2\over H}a^{3/2}\dot\phi
\left[1+2\kappa^2{\phi\dot\phi\over H} +{15\over4}\kappa^2{\dot\phi^2\over m^2}
-{9\over4}{H^2\over m^2}-{\kappa^4\over2}{\dot\phi^4\over m^2H^2}\right]
\nonumber\\
&&=(-1)^j{3\over2}\sigma m^2 C_j\left[m\tau_j \sin x + x\sin x
+{9\over4}(\cos 3x - \cos x) + \Order{x\epsilon_0}\right],
\Eeqr
where
\Beq
x:= m(t-\tau_j),
\Eeq
and $C_j$ is a constant depending on $j$ such that 
\Beq
C_j=1+\Order{\epsilon_0^2},
\Eeq
which is related to the value of $\Tdot{(a^{3/2}\dot\phi/H)}$ at $t=\tau_j$ 
by
\Beq
\Tdot{\left(a^{3/2}\dot\phi\over H\right)}(\tau_j)
=\left({a^{3/2}m^2\phi\over H}\right)(\tau_j)
=(-1)^{j+1}{3\over2}m^2\sigma \tau_j C_j.
\Eeq
The integration of this equation gives the following estimate for
the integrand in \Eq{V2:def}:
\Beq
{a^{3/2}\dot\phi\over H}=(-1)^{j+1}{3\over2}\sigma C_j\sin x
\left[m\tau_j + f(x) + \Order{x^2\epsilon_0}\right],
\Eeq
where
\Beq
f(x):= x -{1\over2}\sin 2x.
\Eeq

Inserting these expressions into the definitions of $V_1$ and $V_2$,
and using the recurrence relation (\ref{RecurrenceRelationForA}), we
obtain
\Beqr
&V_1= & (-1)^{j+1}{3\sigma\over2 a^{3/2}}\left[m\tau_j + f(x)
+\Order{\epsilon_0}\right]\sin x,
\label{V1:estimate:harmonic}\\
&V_2= & (-1)^j{2\over 3m\sigma a^{3/2}}{1\over m\tau_j}
\left(1+{f(x)\over m\tau_j}\right) 
\left[\left(1-{2x\over m\tau_j}\right)\cos x  
+ \Order{\epsilon_0^2}\right].
\label{V2:estimate:harmonic}\Eeqr
From \Eq{xi:estimate} $\sin x$ and $\cos x$ are written in terms of
$t-t_0$ as
\Beqr
&&(-1)^{j+1}\sin x=\cos m(t-t_0) 
-\left({5\over 4mt}-{1\over 4mt_0}\right)\sin m(t-t_0)
+ \Order{\epsilon_0^2},\\
&&(-1)^j\cos x=\sin m(t-t_0) 
+\left({5\over 4mt}-{1\over 4mt_0}\right)\cos m(t-t_0)
+ \Order{\epsilon_0^2}.
\Eeqr
Hence \Eqs{V1:estimate:harmonic} {V2:estimate:harmonic} are written as
\Beqr
&V_1= & {\sqrt{6}\over\kappa}\left[\cos m(t-t_0)
+\left({1\over 4mt_0}-{1\over mt}\right)\sin m(t-t_0)
+{1\over4mt}\sin 3m(t-t_0) +\Order{\epsilon_0^2}
\right]\nonumber\\
&&\label{V1:estimate':harmonic}\\
&V_2= & {4\sqrt{2}\over3\sqrt{3}m\kappa\sigma^2}{1\over (mt)^2}
\left[\sin m(t-t_0)+\left({3\over2mt}-{1\over 4mt_0}\right)\cos m(t-t_0)\right.
\nonumber\\
&&\qquad\qquad\qquad\qquad \left.-{1\over 4mt}\cos 3m(t-t_0) +\Order{\epsilon_0^2}\right].
\label{V2:estimate':harmonic}\Eeqr
These expressions show that in the oscillatory phase we can take $t_j$
so that
\Beqr
&& \sin m(t_j-t_0)=\Order{\epsilon_0},\\
&& V_2(t_j)=0.
\Eeqr
They also show that $V_2$ is approximately in proportion to $\phi/t$,
and takes its local minima and maxima at around $t=\tau_j$. Further
the amplitudes of $V_1$ and $V_2$ are estimated as
\Beqr
&& V_1=\Order{1\over\kappa},\\
&& V_2=\Order{1\over \kappa m^3\sigma^2 t^2}.
\Eeqr
Hence $V_2$ is a decaying mode.

With the help of these estimations we can easily estimate the behavior
of $Y$ in the case in which $k$ is small but nonzero by solving the
integral equation (\ref{IterativeEq:Y}) iteratively. First integrals
appearing in the first-order iteration are calculated as
\Beqr
&& \int_{t_0}aV_1^2dt=
{27\sigma^2\over 40a_0^2 m}(mt_0)^{4/3}(mt)^{5/3} 
\left[1-\left(t_0\over t\right)^{5/3}+{5\over 6mt}\sin 2m(t-t_0) 
+ \Order{\epsilon_0^2}\right],\\
&&\int_{t_0}aV_1V_2 dt =
{7 \over 16 a_0^2m^2} 
\left[1+\Order{\epsilon_0}-\left(t_0\over t\right)^{4/3}
\left\{1-{8\over 7}\sin^2 m(t-t_0)\right\}\right],\\
&& \int_{t_0}aV_2^2dt={2\over21 a^2_0m^4\sigma^2t_0}
\left[1+\Order{\epsilon_0^2}-\left(t_0\over t\right)^{7/3}
\left\{1+ {7\over 6mt}\sin 2m(t-t_0)\right\}\right].
\Eeqr
Hence in the second order in $k$, $Y_1$ and $Y_2$ are given by
\Beqr
& Y_1(t)=&V_1(t)\left[1+{7\over16}{\epsilon_0^2\over l_0^2}
\left\{1 + \Order{\epsilon_0}-\left(t_0\over t\right)^{4/3}
\left(1-{8\over 7}\sin^2 m(t-t_0)\right)\right\}\right]\nonumber\\
&& -{3\sigma mt\over10a^{3/2}}{\epsilon_0\over l_0^2}
\left(t_0\over t\right)^{1/3}\left[1-\left(t_0\over t\right)^{5/3}
+\Order{\epsilon_0}\right]\sin m(t-t_0) + \Order{k^4},
\label{Y1:1st-order:harmonic}\\
& Y_2(t) =&V_2(t)\left[1-{7\over16}{\epsilon_0^2\over l_0^2}
\left\{1 + \Order{\epsilon_0}-\left(t_0\over t\right)^{4/3}
\left(1-{8\over 7}\sin^2 m(t-t_0)\right)\right\}\right],\nonumber\\
&& +{4\over 63}{\epsilon_0\over l_0^2}{1\over m\sigma^2(m t_0)^2}V_1(t)
\left[1-\left(t_0\over t\right)^{7/3}+\Order{\epsilon_0}\right]+\Order{k^4},
\label{Y2:1st-order:harmonic}\Eeqr
where $l$ denotes the ratio of the perturbation wavelength and the
Hubble horizon radius,
\Beq
l:={aH\over k}.
\Eeq

In order to obtain the estimate in full order in $k$, we use an
inequality obtained from \Eq{IterativeEq:Y} instead of the iterative
power series.  Let us consider the integral equation
\Beq
y=v + k^2 G\circ  y.
\Eeq
If we introduce $|y|_m(t)$ and $|v|_m(t)$ defined by
\Beqr
&&|y|_m(t):=\sup_{t_0\le t'\le t}|y(t')|,\\
&&|v|_m(t):=\sup_{t_0\le t'\le t}|v(t')|,
\Eeqr
the integral equation yields the inequality
\Beq
|y(t)|\le |v|_m(t)+k^2\left[|V_2|\int_{t_0}a|V_1|dt
+|V_1|\int_{t_0}a|V_2|dt\right]|y|_m(t).
\Eeq
Since the integrals in the right-hand side of this equations satisfy
the inequalities
\Beqr
&& \int_{t_0}a|V_1|dt\le {9\sigma\over 10a_0^{1/2}m}
(mt_0)^{1/3}(mt)^{5/3}\left[1+\Order{\epsilon_0}\right],\\
&& \int_{t_0}a|V_2|dt\le {2\over a_0^{1/2}m^2\sigma}
{1\over mt_0}\left[1+\Order{\epsilon_0}\right]
\Eeqr
from \Eqs{V1:estimate':harmonic}{V2:estimate':harmonic}, 
we obtain
\Beq
|y(t)|\le |v|_m(t)+{8\over3}{\epsilon_0\over l_0^2}|y|_m(t).
\Eeq
Since $|v|_m(t)$ and $|y|_m(t)$ are nondecreasing functions, we can
replace $|y(t)|$ by $|y|_m(t)$ in this inequality. Hence we obtain the
estimate
\Beq
|y(t)|\le|y|_m(t)\le {|v|_m(t)\over 1-3\epsilon_0 l_0^{-2}}.
\label{Y:Inequality:harmonic}\Eeq

Now let us apply this inequality to $Y_1$ and $Y_2$. First we rewrite
the integral equation for $Y_1$,
\Beq
Y_1=V_1 + k^2G\circ Y_1,
\Eeq
as
\Beq
Y_1-V_1 = k^2 G\circ V_1 + k^2 G\circ(Y_1-V_1),
\Eeq
and apply the inequality (\ref{Y:Inequality:harmonic}) by putting
$y=Y_1-V_1$ and $v=k^2 G\circ V_1$. Then since the first-order
estimate \Eq{Y1:1st-order:harmonic} gives
\Beq
k^2|G\circ V_1|\le {\sqrt{6}\over 5\kappa}{\epsilon_0\over l_0^2}
\left[1+\Order{\epsilon_0}\right],
\Eeq
we obtain the inequality
\Beq
|Y_1-V_1|\le {\sqrt{6}\over 5\kappa}{\epsilon_0 l_0^{-2}[1+\Order{\epsilon_0}]
\over 1-3\epsilon_0 l_0^{-2}}.
\label{Y1:estimate:full-order}\Eeq
Similarly from 
\Beq
k^2|G\circ V_2|\le {2\over21}{\epsilon_0\over l_0^2}
{1\over a_0^{3/2}m^2\sigma t_0}\left[1+\Order{\epsilon_0}\right].
\Eeq
the deviation of $Y_2$ from $V_2$ is bounded as
\Beq
|Y_2-V_2|\le {2\over21}{1\over a_0^{3/2}m^2\sigma t_0}
{\epsilon_0 l_0^{-2}[1+\Order{\epsilon_0}]
\over 1-3\epsilon_0 l_0^{-2}}.
\label{Y2:estimate:full-order}\Eeq

With the help of these estimate we can easily determine the behavior
of the curvature perturbation and the Bardeen parameter.  First the
curvature perturbations $\Phi_1$ and $\Phi_2$ corresponding to $Y_1$
and $Y_2$ are given by
\Beqr
&& \Phi_1 ={3\over 5}\left[1-\left(t_0\over t\right)^{5/3}
+\Order{\epsilon_0}+\Order{\epsilon_0\over l_0^2}\right],
\label{Phi1:harmonic}\\
&& \Phi_2 = {\kappa^2\over 2}{H\over ak^2}
\left[1+\Order{\epsilon_0^2\over l_0^2}\right]
+\Order{{1\over\sigma^2 m}{\epsilon_0^3\over l_0^2}}.
\label{Phi2:harmonic}\Eeqr
Hence $\Phi_2$ decreases in proportion to $t^{-5/3}$ while $\Phi_1$
approaches -3/5 for $t\gg t_0$.

On the other hand the Bardeen parameter is simply related to $Y$ as
$\BP=-HY/\dot\phi$. Hence it oscillatorilly diverges. However, its
value is well-controlled and stays almost constant except for very
small intervals around $t=\tau_j$ in the situation corresponding to
the reheating phase after inflation.

The inflationary phase, during which $\dot\phi$ is very small,
terminates when $\epsilon=H/m$ becomes smaller than unity and the
scalar field begins to oscillate. Since $\BP$ stays constant during
the inflationary stage\CITE{Bardeen.J&Steinhardt&Turner1983}, and
since $|\Phi/\BP|$ is of order $\dot\phi^2/ \rho$ and is much smaller
than unity at the end of the slow-rolling
phase\CITE{Kodama.H&Sasaki1984}, the behavior of the perturbations in
the oscillatory phase after the inflation is well described by the
solution $Y$ with the initial condition $\Phi(\tau_0)=0$ and
$\BP(\tau_0)=\BP_0$, where $\BP_0$ is the value of $\BP$ during the
inflationary stage. From \Eq{BardeenParamByY} and \Eq{PhiByY} this
initial condition is expressed in terms of $Y$ as
\Beq
Y(\tau_0)=0, \quad \dot Y(\tau_0)=-{\ddot\phi_0\over H_0}
\BP_0,
\label{IC:Y:post-inf}\Eeq
where the suffix $0$ denotes the values at $t=\tau_0$.

If we write $Y$ as the linear combination of the fundamental solutions
$Y_1$ and $Y_2$ in the form
\Beq
Y=AY_1 + BY_2,
\Eeq
the constant $A$ is expressed in terms of $\BP_0$ as
\Beq
A=-W(Y_2,Y)={a_0^3\ddot \phi_0\over H_0}\BP_0 Y_2(\tau_0),
\Eeq
since $W(Y_1,Y_2)=W(V_1,V_2)=1$. From \Eq{V2(tau)} and 
\Eq{Y2:estimate:full-order} $Y_2(\tau_0)$ is given by
\Beq
Y_2(\tau_0)=-{H_0\over a_0^3\ddot\phi_0}+
\Order{{1\over a_0^{3/2}\sigma m}{\epsilon_0^2\over l_0^2}}.
\Eeq
Hence we obtain
\Beq
A=-\left[1+\Order{\epsilon_0\over l_0^2}\right]\BP_0.
\Eeq
Similarly from
\Beqr
&& B=W(Y_1,Y)=-{a_0^3\ddot\phi_0\over H_0}\BP_0Y_1(\tau_0),\\
&& Y_1(\tau_0)=\Order{{1\over\kappa}{\epsilon_0\over l_0^2}},
\Eeqr
we obtain
\Beq
B=\Order{\sigma^2m^2\over H_0 l_0^2}\BP_0.
\Eeq
Hence from \Eqs{Phi1:harmonic}{Phi2:harmonic} the behavior of
$\Phi$ is determined as
\Beqr
&\Phi(t) &= A\Phi_1(t) + B\Phi_2(t) \nonumber\\
&&={3\over 5}\BP_0\left[1-\left(t_0\over t\right)^{5/3}
+\Order{\epsilon_0}+\Order{\epsilon_0\over l_0^2}\right].
\Eeqr
Thus $\Phi$ stays constant for $t\gg t_0$. 

On the other hand, in the interval $t_j<t<t_{j+1}$, the Bardeen parameter
$\BP=-HY/\dot\phi$ is given by
\Beq
\BP(t)=\BP_0\left[1+\Order{\epsilon_0\over l_0^2}
+\Order{{1\over \sin m(t-\tau_j)}{\epsilon_0\over l_0^2}
\left(t_0\over t\right)^{1/3}}\right].
\Eeq
Hence at least except for the small intervals $m|t-\tau_j|\le 1/ l_0$,
$\BP$ coincides with the constant $\BP_0$ with the accuracy
$\r{O}(\epsilon_0/l_0)$, which is of order $\e{-20}$ for perturbations
relevant to the present large scale structures.

\subsection{$U=\lambda|\phi|^n/n$ case}

Let us extend the analysis in the previous subsection to a more
general case in which the potential is given by
\Beq
U={\lambda\over n}|\phi|^n,
\Eeq
where $n$ is a positive constant equal to or greater than 2.

First note that from the equations for the background quantities
\Beqr
&& \ddot \phi + 3H\dot \phi + \lambda|\phi|^n/\phi=0,\\
&& H^2={\kappa^2\over 3}\rho={\kappa^2\over3}
\left({\dot\phi^2\over 2}+{\lambda\over n}|\phi|^n\right),
\Eeqr
the condition for $\phi$ to oscillate, $|H\dot\phi|\ll \lambda|\phi|^{n-1}$
is written as
\Beq
\kappa|\phi|\ll 1.
\Eeq
Under this condition, from 
\Beqr
&0&=\phi\ddot\phi+ 3H\phi\dot\phi + \lambda|\phi|^n \nonumber\\
&&=\Tdot{\left(\phi\dot\phi+{3\over2}H\phi^2\right)}
-\left(1-{3\over4}\kappa^2\phi^2\right)\dot\phi^2
+nU(\phi),
\Eeqr
we obtain the cosmic virial theorem
\Beq
<U(\phi)>={1\over n}<\dot\phi^2>\left[1+\Order{\epsilon^2}
+\Order{HT}\right],
\Eeq
where $<Q>$ denote the time average of $Q$ over the cosmic expansion 
time scale, $T$ is the period of oscillation and 
\Beq
\epsilon:=\kappa <\phi^2>^{1/2}
\Eeq
which is of the same order as $\epsilon$ introduced in the previous
subsection for $U=m^2\phi^2/2$.

From this virial theorem $<\rho>$ is expressed in terms of
$<\dot\phi^2>$ as
\Beq
<\rho>={n+2\over 2n}<\dot\phi^2>\left[1+\Order{\epsilon^2}+\Order{HT}\right].
\Eeq
Hence from the energy equation
\Beq
\dot\rho =-3H\dot\phi^2,
\Eeq
we obtain the following evolution equation for $<\rho>$:
\Beq
\Tdot{<\rho>} = - {6n\over n+2}H<\rho>
\left[1+\Order{\epsilon^2}+\Order{HT}\right].
\Eeq
Solving this equation, we obtain 
\Beqr
&& <\rho>\propto a^{-{6n\over n+1}},\\
&& a \propto t^{n+2\over 3n},\\
&& <\phi^2> \propto a^{-{12\over n+2}}.
\Eeqr
To be precise, these equations hold with the accuracy
$\r{O}(\epsilon)$.

Using these equations we can determine the behavior of the
perturbations by the same method as in the previous subsection. First,
since the fundamental solutions $V_1$ and $V_2$ for $k=0$ are bounded
as
\Beqr
&& |V_1|\le {\sqrt{6}\over \kappa},\\
&& |V_2|=\Order{\kappa\epsilon\over Ha^3}\propto t^{-4\over n},
\Eeqr
the integrals of $aV_jV_k$ are calculated as
\Beqr
&& \int_{t_0}a V_1^2dt ={1\over\kappa^2}{9n^2\over (n+2)(2n+1)}
(at-a_0t_0)\left[1+\Order{\epsilon_0}\right],\\
&& \int_{t_0}a|V_1V_2|dt=\Order{\epsilon\over a^2H^2},\\
&&\int_{t_0} aV_2^2 dt =\Order{\kappa^2\epsilon^2\over a^5H^3}.
\Eeqr
Hence the deviation of $Y_j$ from $V_j$ is estimated as
\Beqr
&&|Y_1-V_1|=\Order{{1\over\kappa}{\epsilon\over l^2}}
=\Order{V_1}\Order{\epsilon\over l^2},\\
&&|Y_2-V_2|=\Order{{\kappa\over a^3H}{\epsilon^2\over l^2}}
=\Order{V_2}\Order{\epsilon\over l^2}.
\Eeqr
From this estimate the behavior of $\Phi_j$ are determined as
\Beqr
&&\Phi_1=-{3n\over 2(2n+1)}\left[1-{a_0t_0\over at}+\Order{\epsilon_0}
+\Order{\epsilon\over l^2}\right],\\
&&\Phi_2={\kappa^2 H\over 2ak^2}\left[1+\Order{\epsilon\over l^2}\right]
\propto t^{-{2(2n+1)\over 3n}}.
\Eeqr
Hence $\Phi_2$ gives a decay mode while $\Phi_1$ approaches a constant
for $t\gg t_0$, as in the harmonic case.

In the situation considered at the end of the previous subsection
these estimates lead to the same conclusion. In fact for the solution
$Y$ satisfying the initial condition (\ref{IC:Y:post-inf}), $A$ and
$B$ are given by
\Beqr
&& A=-\BP_0\left[1+\Order{\epsilon_0\over l_0^2}\right],\\
&& B=\Order{a_0^3H_0\over \kappa^2l_0^2}\BP_0.
\Eeqr
Hence we obtain
\Beqr
&& \Phi(t)={3n\over 2(2n+1)}\BP_0 \left[1-{a_0t_0\over at}
+\Order{\epsilon_0}+\Order{\epsilon\over l^2}
+\Order{\epsilon_0\over l_0^2}\right],
\label{Phi:post-inf}\\
&& \BP(t)=\BP_0\left[1+\Order{\epsilon_0\over l_0^2}
+\Order{\epsilon\over l^2}
+\Order{{H_0\over H^2(t-\tau_j)}{\epsilon^2\over l_0^2}}\right],\nonumber\\
&&\label{C:post-inf}\Eeqr
for $t_j<t<t_{j+1}$.  Thus $\BP(t)$ coincides with $\BP_0$ with the
accuracy $\r{O}(\sqrt{\epsilon}/l)$ except for tiny periods
$H|t-\tau_j|<\r{O}(\sqrt{\epsilon}/l)$.

\section{Comparison with the system obtained by WKB approximation}

In this section, after showing that when averaged over a spacetime
region of the Hubble radius scale the behavior of the spacetime
structure and the energy-momentum tensor of a nearly homogeneous
oscillating scalar field on an expanding universe is well described by
a perfect fluid, we compare the behavior of the curvature perturbation
and the Bardeen parameter of that fluid system with those considered
in the previous section.

In order to find the averaged behavior of the energy-momentum tensor
of a scalar field, let us extend the cosmic virial theorem for an
exactly homogeneous scalar field derived in \S3.2 to a nearly
homogeneous case. The argument is quite similar to that there. First
we multiply the field equation
\Beq
\Box \phi -U_\phi=0,
\Eeq
by $\sqrt{-g}\phi$ to get
\Beqr
&0&=\sqrt{-g}\phi\Box\phi -\sqrt{-g}\phi U_\phi\nonumber\\
&&=\partial_\mu(\sqrt{-g}\phi\nabla^\mu\phi)
-\sqrt{-g}\left((\nabla\phi)^2 + \phi U_\phi\right).
\Eeqr
By integrating this equation over a spacetime region of a proper size
$L$, we obtain
\Beq
<\sqrt{-g}(\nabla\phi)^2> + <\sqrt{-g}\phi U_\phi>
=\Order{{1\over L}<\sqrt{-g}|\phi||\nabla\phi|>}.
\Eeq
Hence if the parameter $\epsilon$ defined by
\Beq
\epsilon:={1\over L}{<\sqrt{-g}|\phi||\nabla\phi|>
\over <\sqrt{-g}(\nabla\phi)^2>}
\Eeq
is much smaller than unity, we obtain the following relativistic
virial theorem for the potential $U=\lambda|\phi|^n/n$:
\Beq
<\sqrt{-g}U>=-{1\over n}<\sqrt{-g}(\nabla\phi)^2>[1+\Order{\epsilon}].
\Eeq
Note that $\epsilon$ is of the same order as that in the previous
section if $L$ is taken to be $1/H$.

Applying this virial theorem to the energy-momentum tensor of the
scalar field,
\Beq
T^\mu_\nu=\nabla^\mu\phi\nabla_\nu \phi-\delta^\mu_\nu
\left({1\over2}(\nabla\phi)^2+U\right),
\Eeq
we obtain
\Beq
<\sqrt{-g}T^\mu_\nu>
=<\sqrt{-g}g^{\mu\lambda}\partial_\lambda\phi\partial_\nu\phi>
+{n-2\over 2n}\delta^\mu_\nu<\sqrt{-g}(\nabla\phi)^2>[1+\Order{\epsilon}].
\Eeq
If the scalar field and the spacetime is nearly spatially homogeneous,
we can neglect the correlation among $g_{\mu\nu}$ and $\phi$ up to the
linear order of perturbation:
\Beqr
&&<\sqrt{-g}g^{\mu\lambda}\partial_\lambda\phi\partial_\nu\phi>
\simeq \sqrt{-g}g^{\mu\lambda}<\partial_\lambda\phi\partial_\nu\phi>,\\
&&<\sqrt{-g}T^\mu_\nu>\simeq \sqrt{-g}<T^\mu_\nu>,
\Eeqr
where we have written $<g_{\mu\nu}>$ simply as $g_{\mu\nu}$.  In this
approximation the spacetime average of $T^\mu_\nu$ is given by
\Beq
<T^\mu_\nu>=<\nabla^\mu\phi\nabla_\nu\phi>
+{n-2\over 2n}\delta^\mu_\nu <(\nabla\phi)^2>.
\Eeq

When the scalar field oscillates rapidly enough and the spacetime is
nearly flat on the time scale of oscillation, it can be assumed that
$\phi$ is well approximated by the WKB form
\Beq
\phi=F(S,x),
\Eeq
where $S$ is a rapidly oscillating phase, and $\nabla F$(the partial
derivative of $F$ with respect to the argument $x$) is much smaller
than the norm of the vector $\nabla\phi$:
\Beq
\epsilon^2=\Order{<(\nabla F)^2>\over <(\nabla\phi)^2>}
=\Order{<(\nabla F)^2>\over <|\partial_S F|^2(\nabla S)^2>}
\ll 1.
\Eeq
In this situation the correlation between $(\partial_S F)^2$ and
$\partial_\mu S$ is small:
\Beq
<(\partial_S F)^2\partial_\mu S\partial_\nu S>
=<(\partial_S F)^2>\partial_\mu<S> \partial_\nu<S>
+\Order{\epsilon<(\nabla\phi)^2>}.
\Eeq
Hence if we introduce two scalar functions $\rho$ and $P$, and a
time-like unit vector $U^\mu$ by
\Beqr
&&(\rho+P)U_\mu U_\nu :=<(\partial_S F)^2>\partial_\mu<S> \partial_\nu<S>,\\
&& P:={n-2\over 2n}<(\nabla\phi)^2>,
\Eeqr
the above equation for $<T^\mu_\nu>$ is written as
\Beqr
&&<T^\mu_\nu>\simeq (\rho+P)U^\mu U_\nu + \delta^\mu_\nu P,\\
&& P= w\rho;\quad w={n-2\over n+2}.
\Eeqr
Hence the averaged dynamics of the scalar field is described by a
perfect fluid. In particular when the geometry and the field is
spatially homogeneous, the averaged energy density and the scale
factor behave as
\Beqr
&& \rho \propto a^{-3(1+w)} \propto {1\over t^2},
\label{rhoBya}\\
&& H \propto a^{-{3\over2}(1+w)} \propto {1\over t},
\label{HBya}\\
&& a \propto t^{2\over 3(1+w)}.
\label{aByt}\Eeqr
This behavior coincides with those obtained in the previous section.

Next let us examine the behavior of perturbations of this fluid system
and compare it with that obtained in the previous section. For scalar
adiabatic perturbations for the fluid with $P=w\rho$, their evolution
equations are expressed in terms of the curvature perturbation $\Phi$
and the Bardeen parameter $\BP$ as\CITE{Kodama.H&Sasaki1984}
\Beqr
&& D\Phi + {5+3w\over 2}\Phi={3\over 2}(1+w)\BP,\\
&& D\BP = -{2\over 3l^2}{w\over 1+w}\Phi,
\Eeqr
where $D$ is the differential operator
\Beq
D:=a{d\over da}.
\Eeq
By noting the time-dependence of $l^2$ and $a^3H$ obtained from
\Eq{HBya},
\Beq
{1\over l^2}\propto a^{1+3w},\qquad
a^3H \propto a^{{3\over 2}(1-w)},
\Eeq
this evolution equation leads to the following 2nd-order differential
equation for $\BP$:
\Beq
D(a^3H D\BP)=-w {a^3H\over l^2}\BP.
\Eeq
By the Green function method this equation is converted into the 
integral equation
\Beq
\BP=A + {B\over a^3H} -{2w\over 3(1-w)}\int_{a_0}{\BP\over l^2}{da\over a}
+{2w\over 3(1-w)}{1\over a^3H}\int_{a_0}{a^3H\over l^2}\BP{da\over a}.
\Eeq
From this $\Phi$ is expresses as
\Beq
\Phi={9(1-w^2)\over 4w}{l^2\over a^3H}\left[B
+ {2w\over 3(1-w)}\int_{a_0}{a^3H\over l^2}\BP{da\over a}\right].
\Eeq

Let us denote the solutions to this integral equation as
\Beqr
&&\Phi=A\Phi_1 + B\Phi_2,\\
&&\BP=A\BP_1 + B\BP_2.
\Eeqr
Then by iteration the fundamental solutions $(\Phi_j,\BP_j)$ are
estimated as
\Beqr
&\BP_1 \simeq &1-{2w\over 3(1-w)(1+3w)}\left({1\over l^2}-{1\over l_0^2}\right)
+{4w\over 3(1-w)(5+3w}{1\over l^2}\left(1-{a_0 t_0\over a t}\right),\nonumber\\
&&\\
&\Phi_1 =&{3(1+w)\over 5+3w}\left[1-{a_0t_0\over aH}+
\Order{1\over l^2}\right],\\
&\BP_2 \simeq &{1\over a^3H}\left[1-{4w\over 3(1-w)(9w-1)}{1\over l^2}
+{2w\over 3(1-w)(1+3w)}\left({1\over l^2}-{1\over l_0^2}\right)\right]
\nonumber\\
&&+{4w\over 3(1-w)(9w-1)}{1\over l_0^2a_0^3H_0},\\
&\Phi_2 = & {9(1-w^2)\over 4w}{l^2\over a^3H}\left[1+
\Order{1\over l^2}+\Order{1\over l_0^2}\right].
\Eeqr
These expressions show that $\Phi_j$ for the fluid system with
$w=(n-2)/(n+2)$ behaves exactly in the same manner as those obtained
in the previous section, \Eqs{Phi1:harmonic}{Phi2:harmonic}, apart
from a normalization constant.

A similar conclusion can be obtained for the behavior of the Bardeen
parameter. To see this, let us consider the solution with the initial
condition
\Beq
\Phi(t_0)=0,\quad \BP(t_0)=\BP_0.
\Eeq
Then, since in the present case $\Phi_(t_0)=0$ and $\Phi_2(t_0)\not=0$,
the corresponding solution is proportional to $(\Phi_1,\BP_1)$. Hence 
it is given by
\Beqr
&&\Phi(t)={3n\over 2(2n+1)}\BP_0\left[1-{a_0t_0\over at}
+\Order{1\over l^2}\right],\\
&&\BP(t)=\BP_0\left[1+\Order{1\over l^2}\right].
\Eeqr
These expressions coincide with \Eqs{Phi:post-inf}{C:post-inf} in the
leading order including the coefficients, though the conservation of
the Bardeen parameter holds better in the exact treatment done in the
previous section than that in the WKB fluid approximation.

\section{Summary and discussion}

In this paper we have analyzed the behavior of scalar perturbations on
superhorizon scales in a stage dominated by a rapidly oscillating
scalar field $\phi$ and have shown that the Bardeen parameter stays
constant in a good accuracy expect for a sequence of negligibly short
intervals around the zero points of $\dot \phi$. We have further shown
that the scalar field system reduces to a perfect fluid system by a
spacetime averaging when the WKB approximation is good, and that the
behavior of perturbations for that perfect fluid system coincides in a
good accuracy with that of the original system outside the above
sequence of time intervals.

In realistic situations in which radiation is produced by the decay of
the scalar field (inflaton), the divergence at around $t=\tau_n$ is
replaced by a sequence of narrow peaks and their height decreases as
the energy density of radiation increases.  Hence the above result
practically establishes the conservation of the Bardeen parameter
during the reheating phase dominated by the scalar field, provided
that the energy transfer from the inflaton to radiation does not
affect its behavior.

Further the above result shows that we can study the evolution of
superhorizon perturbations during the reheating phase of inflation
with a good accuracy by replacing the scalar field by a perfect fluid
obtained by the WKB approximation and the spacetime averaging. This
provides a generalization to relativistic superhorizon perturbations
of the result obtained by Nambu and Sasaki in the Newtonian
approximation\CITE{Nambu.Y&Sasaki1990}, and will significantly
simplify the analysis of the evolution of perturbations during
reheating.  The effect of the energy transfer on perturbations during
reheating will be discussed utilizing this simplification in a next
paper.

\section*{Acknowledgments}

The authors thank A. Taruya for valuable discussions. This work is 
partly supported by the Grant-In-Aid for Scientific Research of the 
Ministry of Education, Science, Sports and Culture of Japan(40161947).


\addtolength{\baselineskip}{-3mm}

\end{document}